\documentclass[%
 reprint,
superscriptaddress,
nofootinbib,
 amsmath,amssymb,
 aps,
]{revtex4-2}
\usepackage{graphicx}% Include figure files
\usepackage{dcolumn}% Align table columns on decimal point
\usepackage{bm}% bold math
\usepackage[version=4]{mhchem} %chemical elements
\usepackage{subcaption}
\usepackage{float}
\usepackage{caption}
\usepackage{multirow}
\usepackage{xcolor}
\usepackage[colorlinks,
            linkcolor=red,
            anchorcolor=blue,
            citecolor=blue
           ]{hyperref}% add hypertext capabilities
\graphicspath{ {figure/} }
\begin{document}

\title{An interpretable unsupervised representation learning for high precision measurement in particle physics}
%Slow-Diffusion Disk model}% Force line breaks with \\

\author{Xing-Jian Lv}
%\email{lvxj@ihep.ac.cn}
 \affiliation{%
Institute of High Energy Physics, Chinese Academy of Sciences, Beijing 100049, China}
\affiliation{
 University of Chinese Academy of Sciences, Beijing 100049, China 
}%
 \author{De-Xing Miao}
 \email{miaodx@ihep.ac.cn}
\affiliation{%
 Institute of High Energy Physics, Chinese Academy of Sciences, Beijing 100049, China}
\affiliation{
 University of Chinese Academy of Sciences, Beijing 100049, China 
}%
\author{Zi-Jun Xu}

\affiliation{%
Institute of High Energy Physics, Chinese Academy of Sciences, Beijing 100049, China}
 \author{Jian-Chun Wang}
\affiliation{%
Institute of High Energy Physics, Chinese Academy of Sciences, Beijing 100049, China}

%\affiliation{
%\affiliation{%


\date{\today}% It is always \today, today,

\begin{abstract}
Unsupervised learning has been widely applied to various tasks in particle physics. However, existing models lack precise control over their learned representations, limiting physical interpretability and hindering their use for accurate measurements. We propose the Histogram AutoEncoder (HistoAE), an unsupervised representation learning network featuring a custom histogram-based loss that enforces a physically structured latent space. Applied to silicon microstrip detectors, HistoAE learns an interpretable two-dimensional latent space corresponding to the particle’s charge and impact position. After simple post-processing, it achieves a charge resolution of $0.25\,e$ and a position resolution of $3\, \mu$m on beam-test data, comparable to the conventional approach. These results demonstrate that unsupervised deep learning models can enable physically meaningful and quantitatively precise measurements. Moreover, the generative capacity of HistoAE enables straightforward extensions to fast detector simulations.
\end{abstract}

\maketitle

%\tableofcontents

\section{\label{sec:level1}INTRODUCTION}
Machine learning, and in particular its modern incarnation of deep learning (DL)~\cite{LeCun:2015pmr, Schmidhuber_2015}, has become an indispensable tool in particle physics, a field that routinely handles vast datasets and nonlinear relationships among observables~\cite{Lonnblad:1990bi, Denby:1987rk, PDG2023_ML, Carleo:2019ptp}. In recent years, advances in DL have expanded the scope of data-driven progress across the energy, intensity, accelerator, and cosmic frontiers~\cite{Radovic:2018dip, Guest:2018yhq}. Despite remarkable advancements, most current DL applications in particle physics are supervised, relying either on Monte Carlo (MC) simulations or on labeled experimental data. However, because simulations cannot fully capture the complexity of the real world, a persistent gap between MC and Data leads to training bias. Direct training on real data, in turn, demands truth labels for every event. In practice, such labels must be derived from auxiliary detectors, whose raw measurements cannot serve directly as training targets but require elaborate, detector-specific calibration and manual processing to convert into usable labels~\cite{Yu2025HERD_MLP, Ambrosi:2017gez, Jia:2020zuo}. This labeling pipeline is difficult to generalize and scale. For this reason, the development of unsupervised DL~\cite{bengio2014representationlearningreviewnew, jing2019selfsupervisedvisualfeaturelearning} is integral for particle physics.

Unsupervised learning has achieved remarkable success in tasks such as clustering and identifying topological phases of matter~\cite{Rodriguez-Nieva:2018cbl, Scheurer:2020cii}, anomaly detection~\cite{Kottmann:2020uxr, Cheng:2024yig}, and learning representations that capture essential physical degrees of freedom~\cite{Iten:2020ohp, Hou2024UnsupervisedKS}. These results demonstrate that unsupervised methods can discover physically meaningful structures from raw data without explicit labels.
However, measurements in particle physics demand quantitative precision and interpretability. To date, unsupervised models lack explicit control over their learned representations~\cite{KolouriPopeMartinRohde2019SWAE}, rendering them unsuitable for direct physical measurement~\cite{Fraser:2021lxm}. 
% Therefore, no fully unsupervised DL method has yet demonstrated the ability to achieve precise, physically interpretable measurements.

In this work, we present a fully unsupervised DL approach capable of performing precise measurements. We focus on the charge and position measurements of particles (nuclei) using silicon microstrip detectors (SSDs)~\cite{Seidel:2019hty,Wei:2020lpl}, which are among the most fundamental physical observables in many areas of particle physics, such as cosmic-ray composition and rigidity measurements in spaceborne experiments~\cite{AMS:2021nhj}, and final-state particle identification and track reconstruction in heavy-ion experiments~\cite{Zhou:2022pxl}.

The detector response exhibits a complex and nonlinear dependence on the charge and impact position of the incident particle. Consequently, conventional charge and position reconstruction methods rely on elaborate calibration procedures and prior knowledge of particle species~\cite{Ambrosi:2017gez, Jia:2020zuo}. Such priors are typically derived from auxiliary detectors, introducing considerable detection inefficiency and potential bias. In particular, conventional charge reconstruction requires prior species identification from other subdetectors, creating a circular dependency between charge measurement and particle identification. Moreover, the calibration itself relies on simulation-derived correction functions, so any mismodeling of the detector response propagates directly into the measured charge. An unsupervised method that learns directly from the raw detector signals removes both dependencies, enabling standalone reconstruction without auxiliary detectors or simulation-tuned calibrations. Furthermore, no conventional SSD reconstruction algorithm simultaneously determines both charge and position, because the respective calibrations are performed independently with different auxiliary inputs. While supervised multi-task neural networks can in principle perform such joint regression~\cite{Pata:2021oez}, they require labeled training data derived from simulation or auxiliary detectors.

AutoEncoders (AE)~\cite{autoencoder, BALDI198953} are a widely used framework for unsupervised learning. An AE consists of an encoder that compresses input data into a latent representation and a decoder that reconstructs the original features. By training to minimize
reconstruction error, AEs learn compact representations of the data. However, standard AEs lack explicit control over the latent distribution, rendering the learned representations not fully interpretable and thus unsuitable for precise measurements. Previous extensions such as Variational AutoEncoders (VAEs)~\cite{kingma2014auto,rezende2014stochastic} and Wasserstein AutoEncoders (WAEs)~\cite{tolstikhin2018wasserstein} incorporate regularization terms to impose global constraints on latent distributions, thereby improving overall controllability. Still, there is a lack of fine-grained control over the internal structure of the latent space.

To overcome this limitation, we introduce a novel network architecture, the Histogram AutoEncoder (HistoAE), which employs a custom loss function based on the histogram of the latent space. HistoAE enables explicit control over the latent-space distribution, leading to a fully interpretable latent representation. With this network, we achieve highly precise, unsupervised, and simultaneous reconstruction of particle charge and position.

These results demonstrate that a deep learning model can deliver physically meaningful and quantitatively accurate measurements in a completely unsupervised setting. Beyond its application to SSD, our approach provides a general framework for interpretable, label-free analysis of high-dimensional data.

This article is organized as follows. Section~\ref{sec data} introduces the basics of SSDs and the training and testing data set for this work. Section~\ref{sec methology} presents the HistoAE framework in detail. Section~\ref{sec results} reports the main results, and Section~\ref{sec conclusion} provides a summary.

\section{Data Sample}\label{sec data}
{\centering\textbf{Silicon Microstrip Detector for Particle Measurement}}\newline
SSDs are extensively used for precise position and charge measurements of ionizing particles across several fields in collider experiments~\cite{LHCb:2008vvz,CMS:2008xjf,ATLAS:2008xda}, nuclear physics~\cite{He:2023svm}, as well as in many spaceborne experiments ~\cite{PAMELA:2004seq,Dong:2015qma,Vievering:2020wqk} including Alpha Magnetic Spectrometer (AMS-02)~\cite{Lubelsmeyer:2011zz,Duranti:2013qfz}. The SSD used in this work was designed for the Layer-0 tracking detector upgrade of the AMS-02 experiment, aiming to triple the statistical sample and enhance the nuclei identification capability~\cite{Miao:2025ldv}. The Layer-0 detector is scheduled to be launched to the International Space Station in 2027.

\begin{figure}[htbp]
    \centering
    
    \begin{subfigure}[b]{0.45\textwidth}        \includegraphics[width=\textwidth]{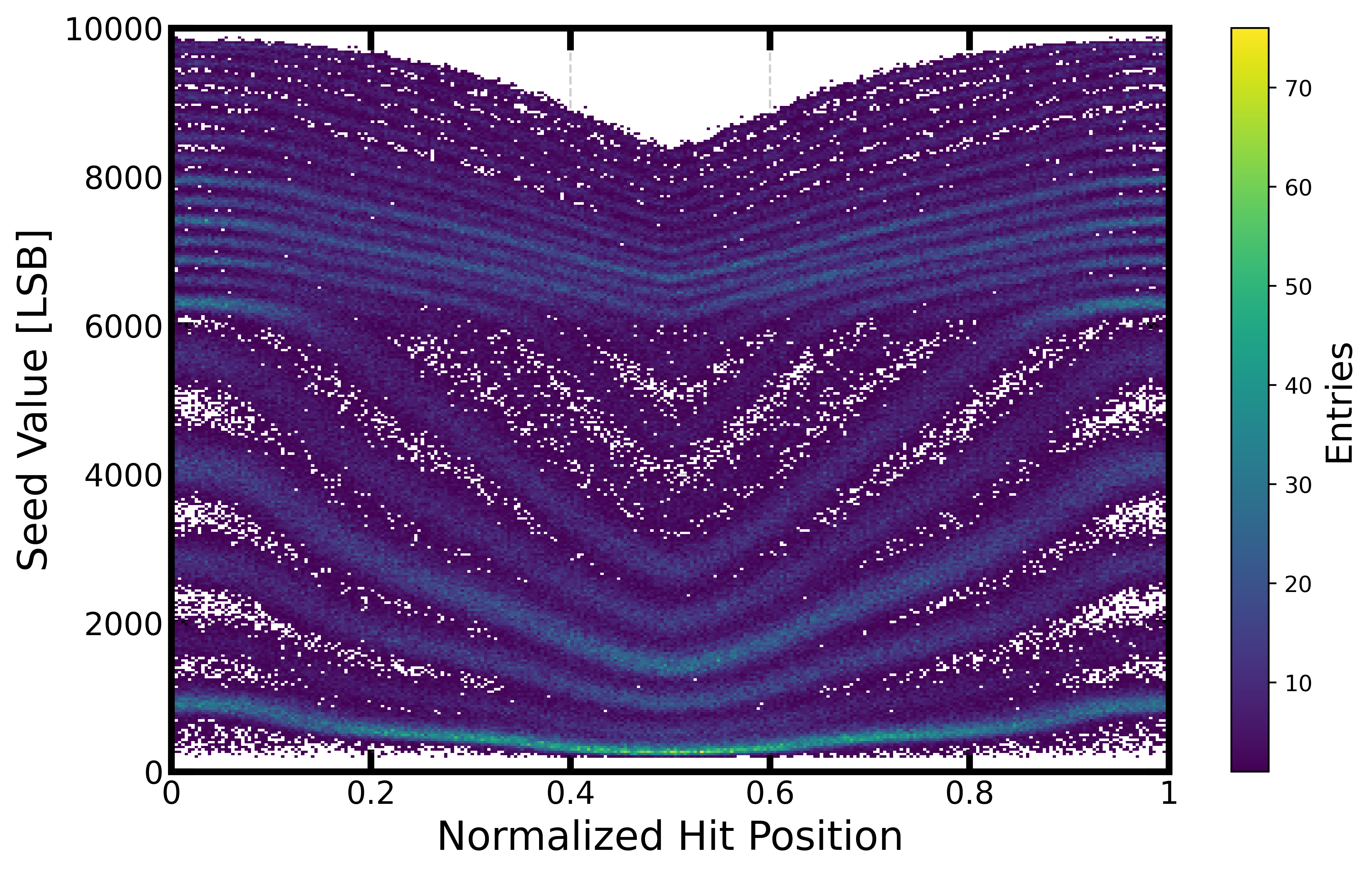}
        %\caption{}
    \end{subfigure}

    \captionsetup{justification=raggedright}
    \caption{Distribution of the maximum channel signal within a cluster (referred to as the seed value) as a function of the normalized impact position for mixed-nuclei beam-test events. The x-axis represents the impact position normalized to the pitch between two readout strips, where a value of 0.5 denotes the midpoint between adjacent strips. The y-axis gives the maximum channel signal in units of the digitized electronics readout (LSB, least significant bit). The horizontal bands from bottom to top correspond to different nuclear species, with the lowest visible band corresponding to $Z=3$ after Hydrogen and Helium events are removed.}
    \label{fig:seed_eta}
\end{figure}

The SSD consists of a large number of parallel, equally spaced readout strips fabricated on a silicon substrate. When a charged particle traverses the detector, it deposits energy along its path through ionization, generating electron–hole pairs in the silicon bulk. Under an applied electric field, charge carriers drift toward the nearest readout strips, causing the signals of several adjacent channels to exceed the threshold. These adjacent channels collectively form what is referred to as a \textit{cluster}. According to the Bethe–Bloch formula~\cite{Bethe1930}, the total deposited energy is proportional to the square of the particle's charge, $Z^2$. However, the signal amplitudes of individual channels within a cluster exhibit a much more complex behavior.

First, due to the nonlinearity and saturation effects of the readout electronics, the signal amplitude recorded by each channel is not strictly proportional to the amount of charge collected by the corresponding readout strip. Moreover, the charge-collection efficiency of the SSD depends on the impact position of the particle: when the particle hits closer to a readout strip, the efficiency is higher. Consequently, even for the same particle species, clusters produced at different impact positions can exhibit different total amplitudes and shapes. In addition, the presence of delta rays leads to a non-Gaussian signal distribution for individual particles~\cite{Boronat2014_PhysicalLimitationsSiliconDetectors}. Fig.~\ref{fig:seed_eta} shows that the signal amplitude of the maximum channel within a cluster, hereafter referred to as the seed value, varies with the particle’s impact position, normalized to the interval between two readout strips. The signal is expressed in LSB, the least significant bit of the digitized electronics readout. Each band corresponds to a specific species; in the figure shown here, the lowest visible band starts at $Z=3$ because Hydrogen and Helium events are removed. The shapes of these bands differ significantly and exhibit complex nonlinear behavior. This complexity makes it challenging to reconstruct the particle’s charge and impact position from the cluster signals.

{\centering\textbf{High Energy Nuclei Beam Test}}\newline
In this work, all training and testing data are real data obtained from beam test measurements, which were carried out at CERN Super Proton Synchrotron (SPS). The beam was generated by bombarding a Lead primary beam of $150\, \mathrm{GeV/n}$ on a Beryllium target. The resulting secondary fragmented beam consisted of a mixture of various nuclei and was selected by the magnetic optics with a mass-to-charge ratio of $A/Z = 2$, while the beam momentum was controlled by the magnet system with a precision of $0.1\%$~\cite{Altuna:1992SPSMomentum}. A total of 5~M data samples were collected for training and testing.

We have a total of nine SSD layers with identical designs. Among them, eight layers serve as a beam telescope, providing reference measurements of the particle charge and impact position, while the central layer is the detector under test (DUT). The “truth” position used in this study refers to the predicted impact point derived from the telescope reconstructed track. The “truth” charge is defined as the average charge value obtained from independent measurements of the same particle track in different telescope layers. The results presented for the network training and testing are the DUT’s completely independent measurements of the same particle tracks.
We emphasize that the ``truth'' charge and position from the telescope are used exclusively for post-training evaluation and are never provided to the network during training. The HistoAE model is trained in a purely unsupervised manner using only the raw DUT cluster signals as both input and reconstruction target.

% \newpage

\section{Model Architecture}\label{sec methology}
\begin{figure}[htbp] 
% \begin{figure}[h] 
\centering 
\includegraphics[width=0.9\hsize]{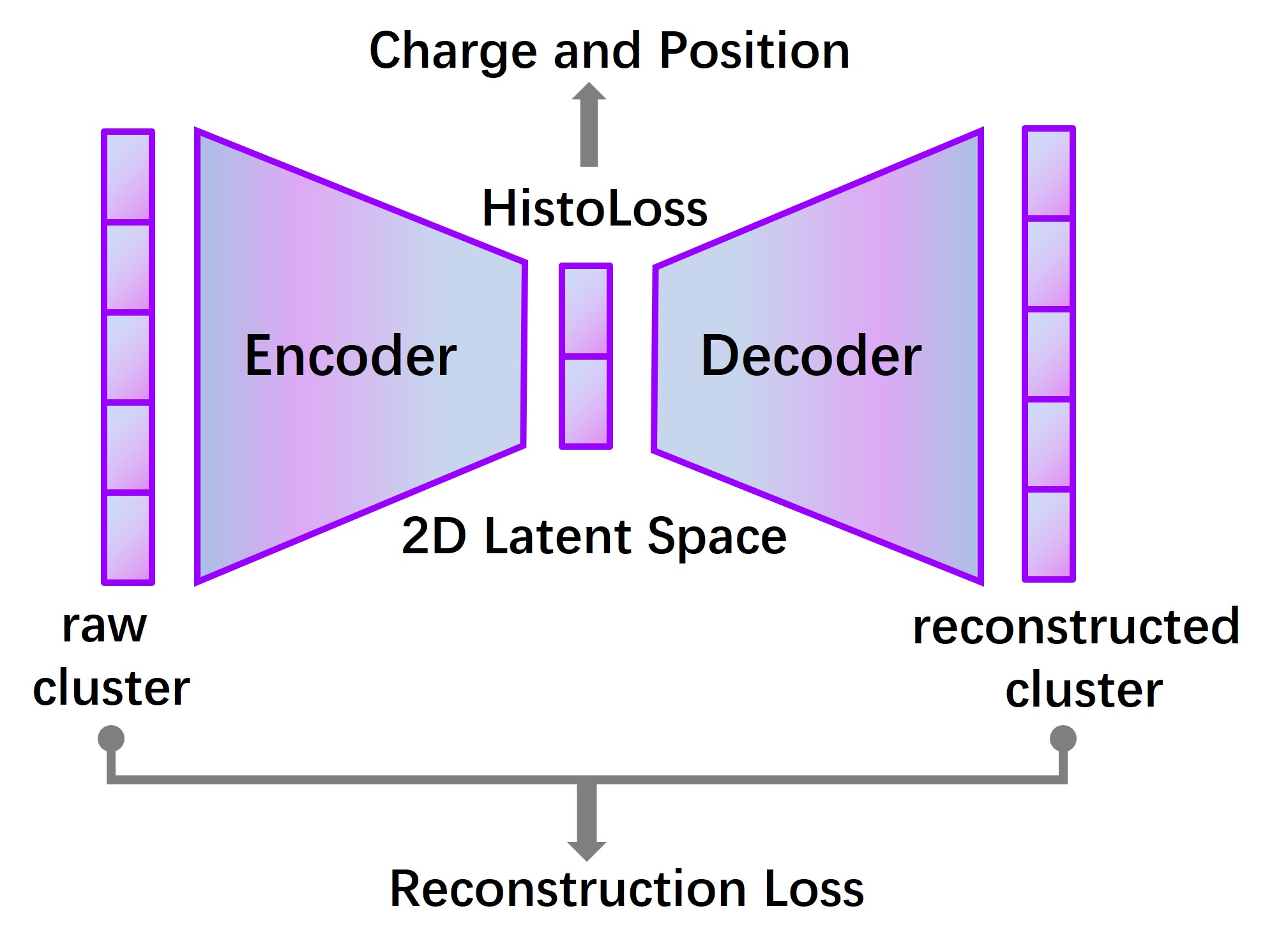} 
\captionsetup{justification=raggedright}
\caption{The network architecture of HistoAE. It consists of an encoder and a decoder connected through a two-dimensional latent space. A reconstruction loss ensures that the reconstructed clusters closely match the original ones, while the HistoLoss constrains the two latent dimensions to represent charge and position, respectively.
\label{fig:HistoAE}}
\end{figure}

{\centering\textbf{Network Design}}\newline
HistoAE adopts a deep neural network (DNN) architecture composed of fully connected layers for both the encoder and decoder. The model is designed to learn compact, physically interpretable latent representations of detector responses, corresponding to the impact position and charge of incident particles. The architecture of HistoAE is shown in Fig.~\ref{fig:HistoAE}.

\begin{figure*}[htbp]
    \centering
    \begin{subfigure}[b]{0.45\textwidth}
        \includegraphics[width=\textwidth]{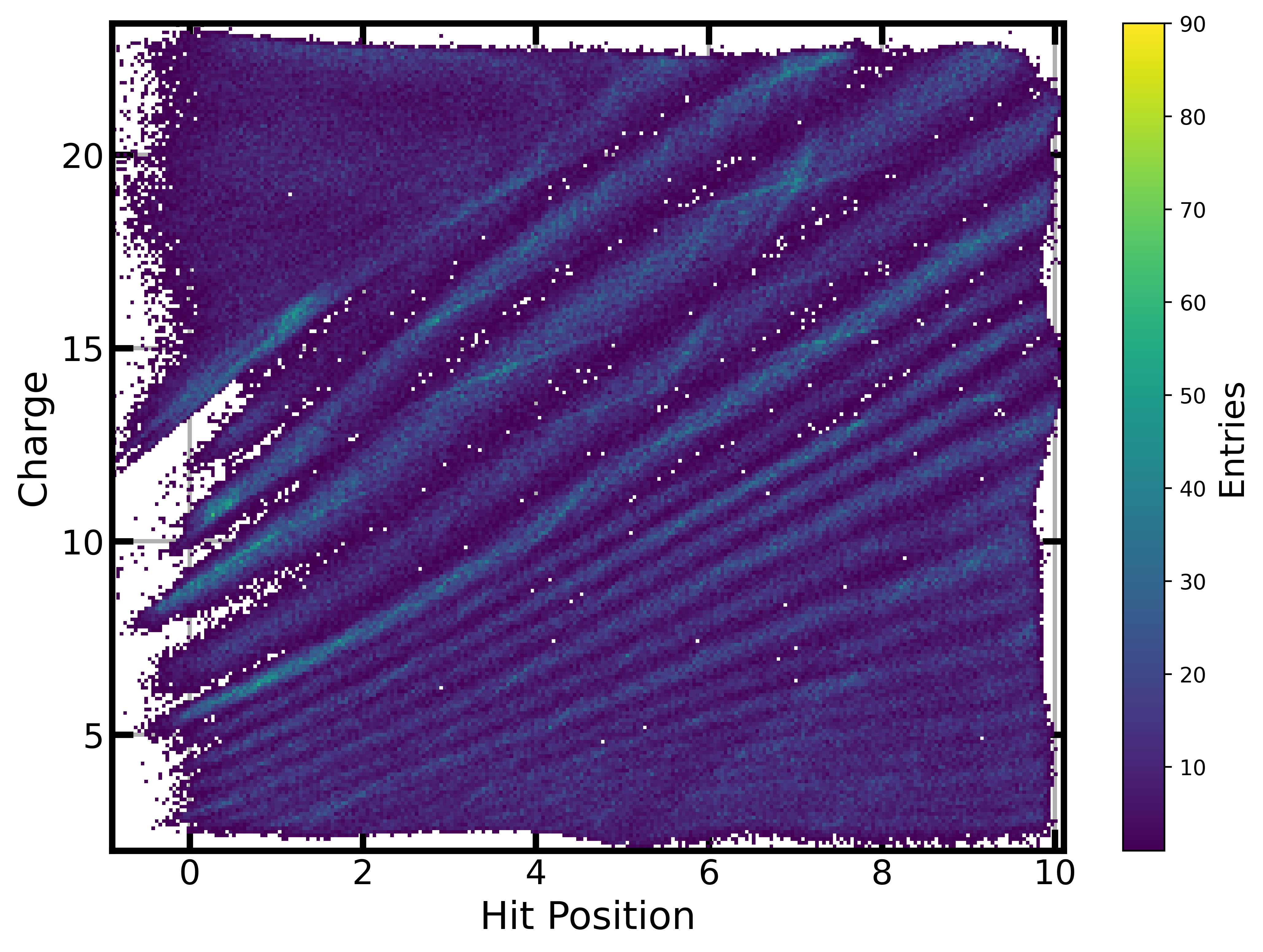}
        \caption{}
    \end{subfigure}
    \begin{subfigure}[b]{0.45\textwidth}
        \includegraphics[width=\textwidth]{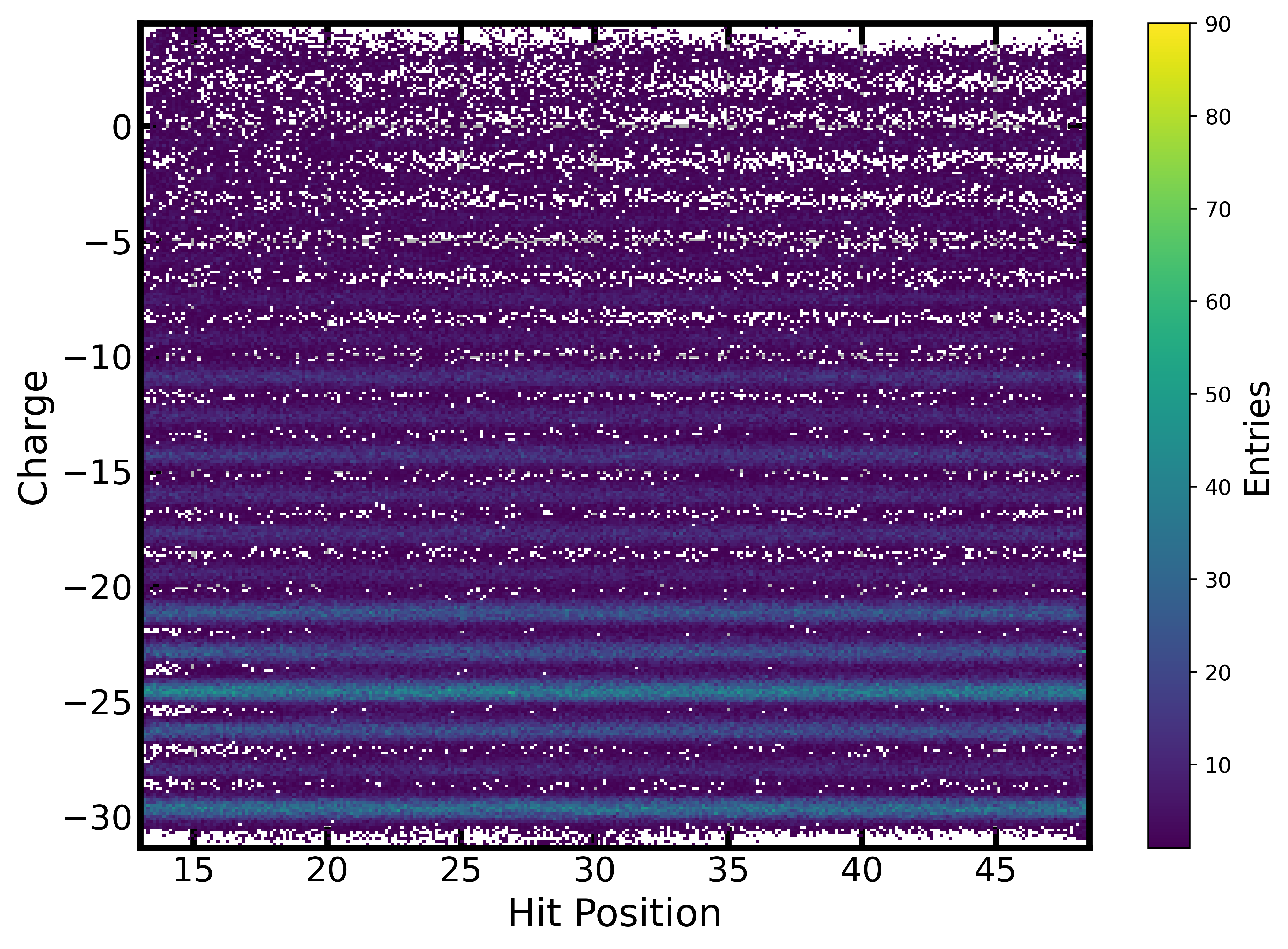}
        \caption{}
    \end{subfigure}

    \vskip 0.3cm

    \begin{subfigure}[b]{0.45\textwidth}
        \includegraphics[width=\textwidth]{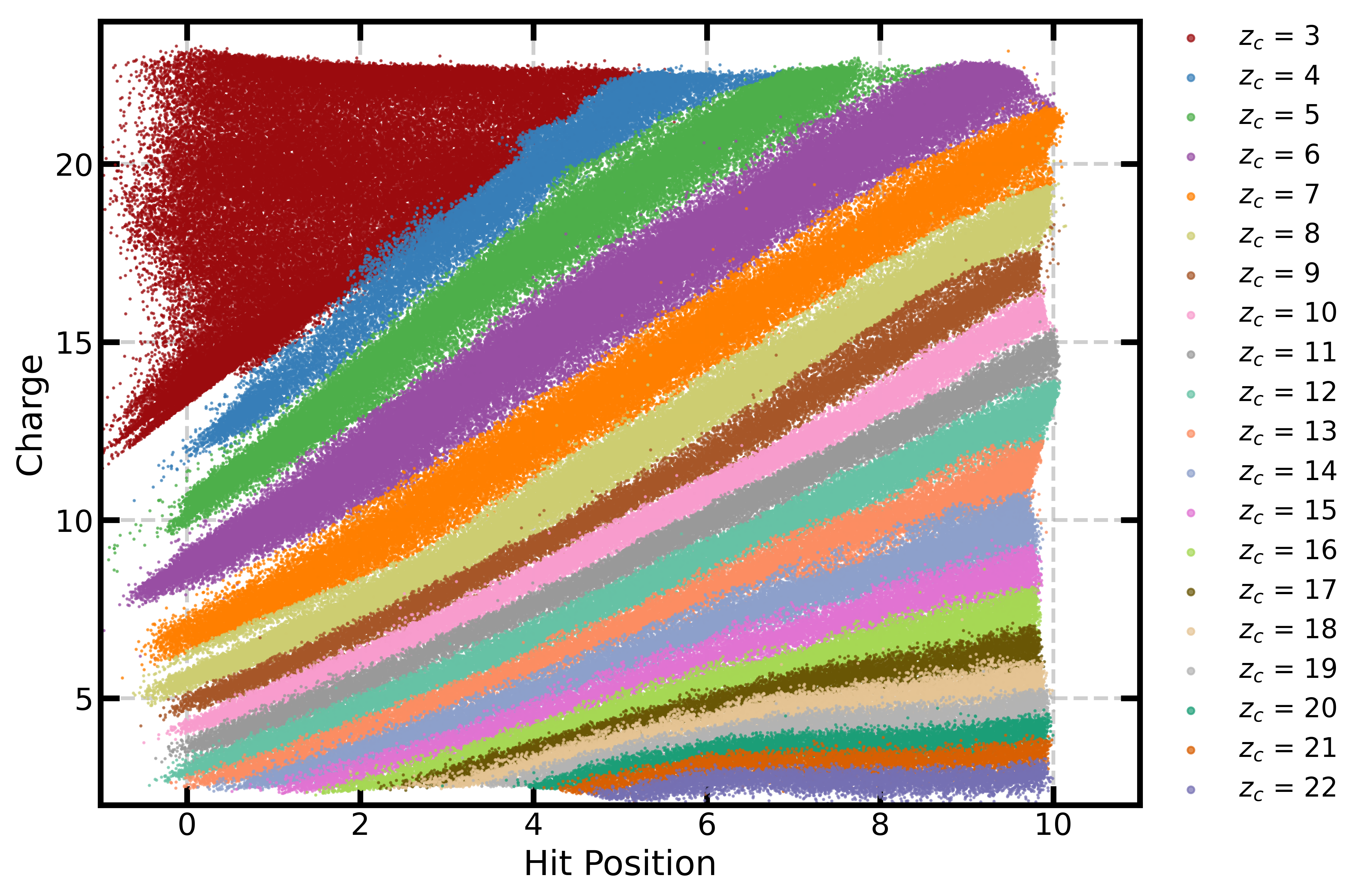}
        \caption{}
    \end{subfigure}
    \begin{subfigure}[b]{0.45\textwidth}
        \includegraphics[width=\textwidth]{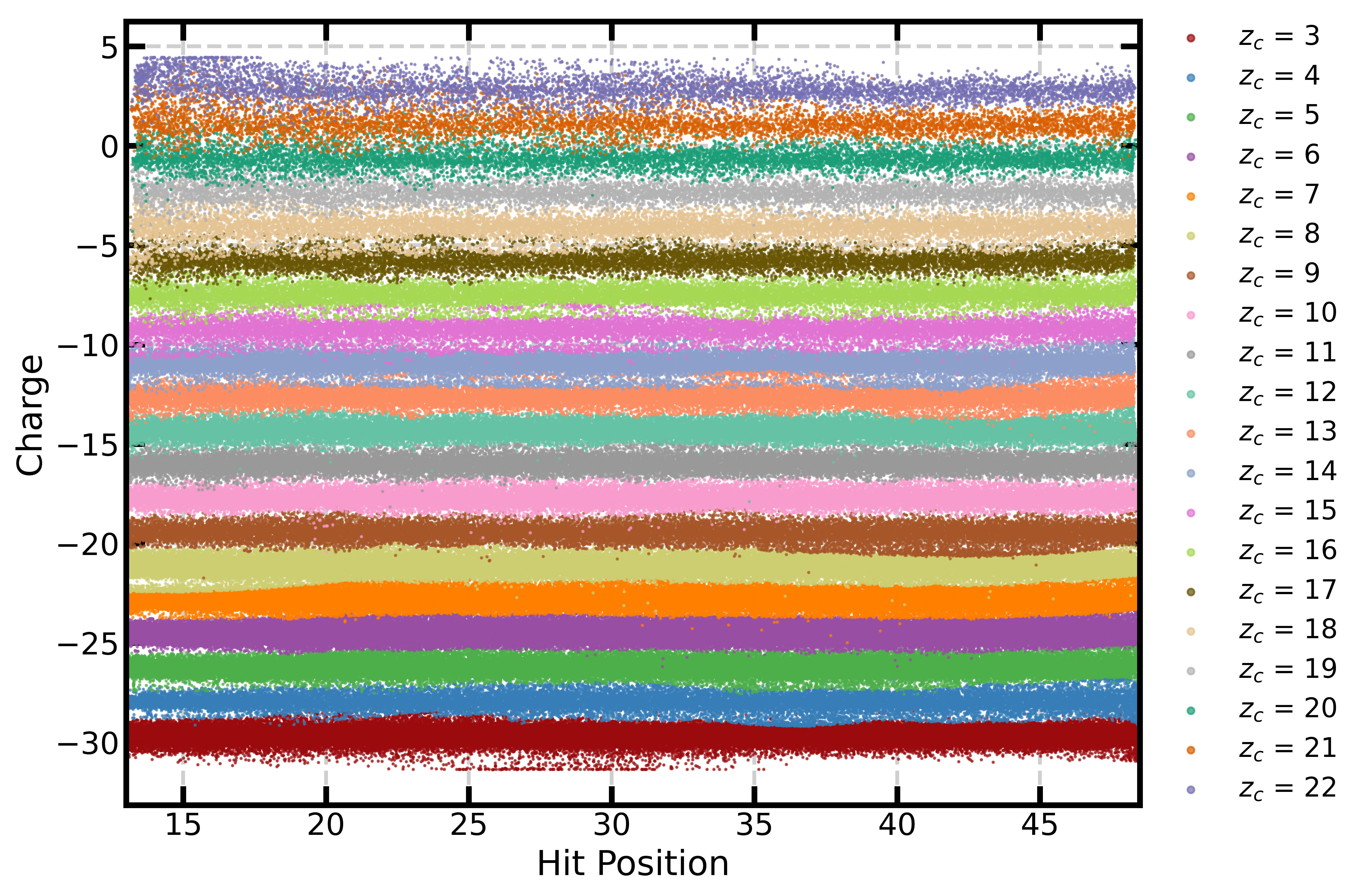}
        \caption{}
    \end{subfigure}

    \captionsetup{justification=raggedright}
    \caption{Latent-space visualizations. (a) and (b) show two-dimensional histograms of the latent variables obtained with the WAE and HistoAE, respectively. (c) and (d) present the corresponding latent spaces color-coded by the true nuclear charge, revealing overlapping diagonal bands for the WAE and well-separated horizontal parallel structures for the HistoAE.}
    \label{fig:2dlatent}
\end{figure*}

\paragraph{Input representation.}  
For each event, we use the readout signals of the five channels with the highest amplitudes within the cluster associated with the incident particle. The raw signals span several orders of magnitude—from $\mathcal{O}(1)$ to $\mathcal{O}(10^4)$—owing to variations in impact position and particle charge. Direct application of standard normalization techniques, such as min–max or mean–variance scaling, fails to alleviate this dynamic range problem, since they preserve the multiplicative scale differences between large and small signals. Logarithmic or power transformations can compress the range but distort the absolute differences between signal amplitudes from adjacent nuclei, thereby hindering the learning of relative charge patterns.

To address this, we introduce a digit-wise vector encoding scheme, referred to as vecoding. Each scalar input value is decomposed into its individual decimal digits and mapped to a fixed-length vector representation. For instance, a value of $9764.4$ is transformed into the vector $[0,\,9,\,7,\,6,\,4,\,4]$, corresponding to a six-dimensional encoding with one decimal place retained. This approach ensures that every element entering the network lies within the range $[0, 9]$, thereby maintaining numerical stability while preserving the intrinsic structure of the original signals. Crucially, this transformation does not alter the relative or absolute information content of the input data. Related digit-level encoding strategies have appeared in machine-learning models for mathematical reasoning~\cite{Charton2023GCD} and in continuous numerical tokenization for scientific data~\cite{Golkar2023xVal}; our vecoding is motivated by the specific dynamic-range requirements of physical detector signals.

\paragraph{Encoder and decoder design.}  
The input vectors from the five channels are flattened into a single 30-dimensional vector (5 × 6) before being fed into the encoder. The encoder consists of a stack of fully connected layers with intermediate dimensions of 256, 128, 64, and 32, each followed by batch normalization and SiLU activation functions. The final encoder layer projects the high-dimensional input into a two-dimensional latent space, representing the particle’s impact position and charge. The decoder mirrors the encoder with hidden widths 32, 64, 128, and 256 in reverse order to reconstruct the input readout patterns from the latent representation. The use of batch normalization in all hidden layers stabilizes training and improves convergence in the presence of large dynamic ranges~\cite{Ioffe2015BatchNormalization}.

{\centering\textbf{HistoLoss}}\newline
Physically, a cluster is fully determined by the impact position and the particle charge. In principle, any linear or nonlinear combination of these two variables would allow the decoder to reconstruct the observed cluster. However, only a properly disentangled latent representation can be interpreted in physically meaningful terms. Therefore, a central objective of our approach is to achieve accurate control of the latent-space distribution.

In the broader machine learning community, several frameworks have been proposed to impose global constraints on the latent distribution, such as VAE and WAE. However, when trained on our data, the resulting latent distributions typically exhibit band-like structures—each band corresponding to a distinct particle species—but these bands are often curved and irregular, lacking clear physical interpretation along either latent dimension as shown in Fig.~\ref{fig:2dlatent} (a)(c). This arises because such models regularize only the overall support of the latent space~\cite{KolouriPopeMartinRohde2019SWAE}, without providing the fine-grained control needed for quantitative measurements in our problem, where signal distributions from different particle species intrinsically overlap due to finite resolution.

To explicitly regulate the geometry of the latent space, we introduce a loss function based directly on a two-dimensional histogram, referred to as HistoLoss. A target latent distribution is first defined, and the histogram of the learned latent variables is computed. The L1 loss between the two histograms is then minimized to align the learned distribution with the target.

In our case, the incident particles carry integer charges. After accounting for detector resolution, the charge dimension is expected to follow a Gaussian Mixture Model (GMM) distribution. The incident position, normalized to lie between two adjacent channels, can be regarded as uniformly distributed. Since the charge and position are independent, the target latent distribution can therefore be expressed as a function of the charge coordinate $q$ and the position coordinate $x$:
\[
H_{\text{target}}(q,x)
   = \sum_{k=1}^{K} w_k
     \,\mathcal{N}(q\,|\,\mu_k,\,\sigma_k^2)
     \,\frac{1}{x_{\max}-x_{\min}},
\]
where $\mu_k$, $\sigma_k^2$, and $w_k$ are the mean, variance, and normalized weight of each charge component, respectively. $H_{\text{target}}(q,x)$ represents the target probability density at charge coordinate $q$ and position coordinate $x$.
We stress that $H_{\text{target}}$ is constructed entirely from generic physical priors: incident particles carry integer charges, which after detector-resolution broadening produce a Gaussian mixture, and the normalized inter-strip impact position is approximately uniform. Here, $K$ denotes the number of charge components retained in the selected beam sample, $w_k$ represents their relative abundances in the mixed beam. The parameters $\mu_k$ correspond to the integer charge values. The $\sigma_k$ parameters provide an estimate of the intrinsic detector resolution, which is reflected in the relative widths of the bands in Fig.~\ref{fig:seed_eta}. These parameter values are used only for initialization. After 5000 epochs, they are released and allowed to be learned by the model, enabling them to adapt to the characteristics of the data itself.

Denote $z_q$ and $z_x$ as the latent variables corresponding to charge and position, respectively. For the latent-space distribution, let $\{(z_{q,i}, z_{x,i})\}_{i=1}^{N}$ represent the latent variables for $N$ events in a minibatch.
We first normalize $z_{q,i}$ and $z_{x,i}$ to the range $[0,1]$ and compute their two-dimensional histogram. To enable backpropagation, the histogram is smoothed using a Gaussian kernel, yielding the empirical latent density $H_{\text{emp}}(q,x)$:
\[
H_{\text{emp}}(q,x)
   = \frac{1}{N}
     \sum_{i=1}^{N}
     \mathcal{K}_\sigma(q-z_{q,i})\,
     \mathcal{K}_\sigma(x-z_{x,i}),
\]
where $\mathcal{K}_\sigma$ is a Gaussian kernel with bandwidth $\sigma$.  

The histogram loss is computed as the element-wise $L_1$ distance between the empirical and target densities,
\[
\mathcal{L}_{\text{histo}}
   = \big\|
       H_{\text{emp}}(q,x) - H_{\text{target}}(q,x)
     \big\|_1,
\]
evaluated on the binned latent grid.

The total training objective combines this histogram loss with the standard reconstruction loss,
\[
\mathcal{L}_{\text{total}}
   = \mathcal{L}_{\text{recon}}
     + \lambda_{\text{histo}}\,
       \mathcal{L}_{\text{histo}},
\]
where $\lambda_{\text{histo}}$ controls the strength of the latent-space regularization.

{\centering\textbf{Training Details}}\newline
All model implementation and training were performed using the \texttt{PyTorch} framework~\cite{Paszke2019PyTorch}. The network was trained end-to-end using the AdamW optimizer~\cite{Loshchilov2019AdamW} with a weight decay of $10^{-4}$ and a base learning rate of $10^{-3}$. All weights were initialized following the Kaiming scheme~\cite{He2015Delving} to ensure stable convergence. A linear warm-up~\cite{Goyal2017Accurate} of the learning rate was applied during the first 10 epochs, followed by a cosine annealing schedule~\cite{Loshchilov2017SGDR} that gradually decreased the rate to $10^{-5}$. Mixed-precision training with automatic gradient scaling (\texttt{torch.amp.GradScaler}) was employed to enhance numerical stability and computational efficiency on GPUs.

The dataset was randomly split into 80\% for training and 20\% for testing. Training was conducted for up to 20\,000 epochs. To improve convergence and ensure physical consistency of the latent representation, we adopted a two-stage training strategy. In the first stage, the network was pre-trained using data with charge number $3\le Z\le13$ and a batch size of 10\,000. In the second stage, the model was post-trained with an enlarged batch size of 40\,000 and data covering the range $3\le Z\le22$. The use of a large batch size is particularly important because the number of high-$Z$ events in the beam-test dataset is relatively small. Smaller batches would contain too few samples of these rare species to yield stable gradients or meaningful statistics for the empirical latent histogram $H_{\text{emp}}(q,x)$ used in the HistoLoss. We start from $Z = 3$ for the same reason, since Hydrogen and Helium dominate the statistics and can be easily selected with a simple value cut.

\section{RESULTS}\label{sec results}

\begin{figure}[htbp]
    \centering
    \begin{subfigure}[b]{0.48\hsize}
        \includegraphics[width=\textwidth]{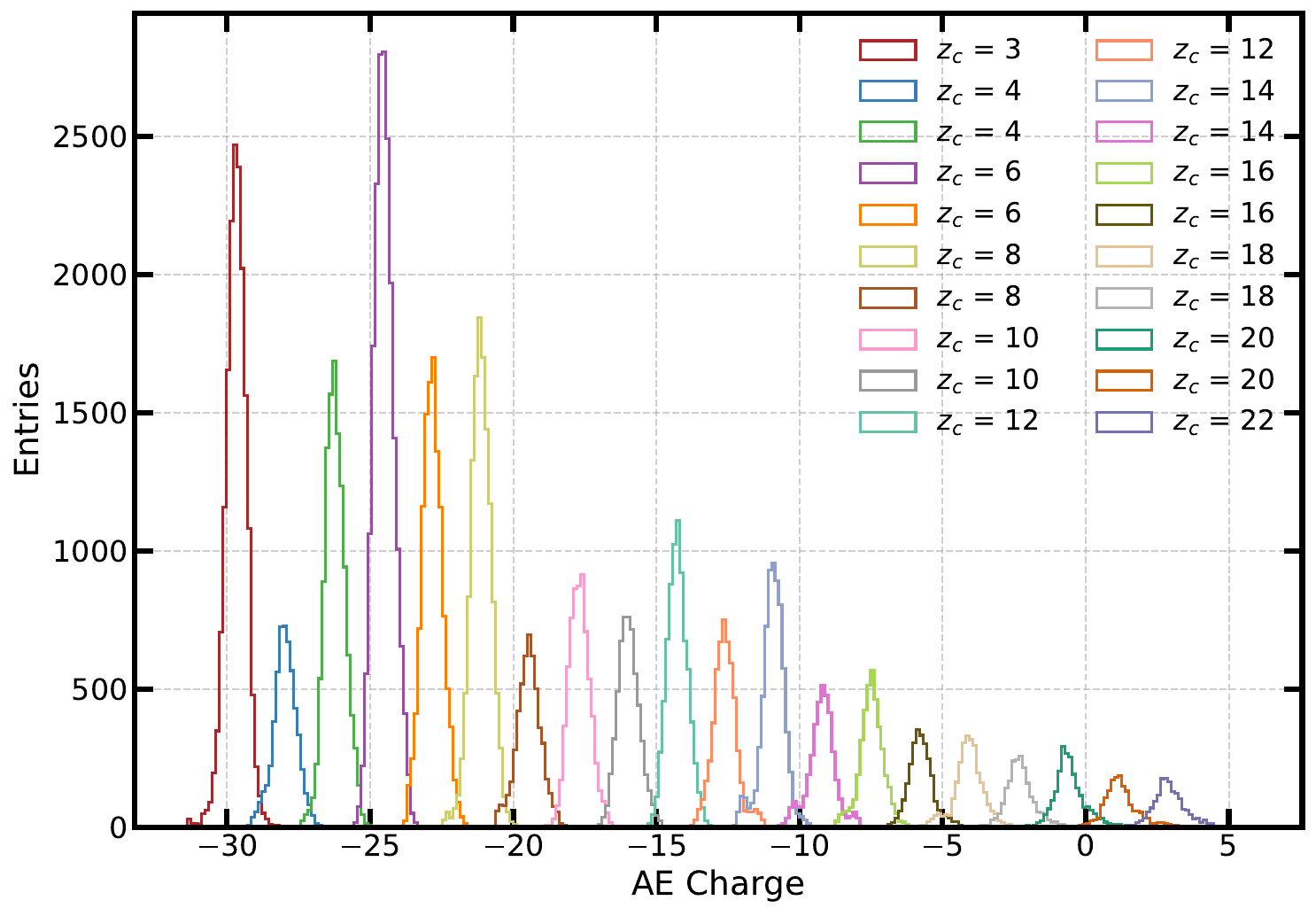}
        \caption{}
    \end{subfigure}
    \begin{subfigure}[b]{0.45\hsize}
        \includegraphics[width=\textwidth]{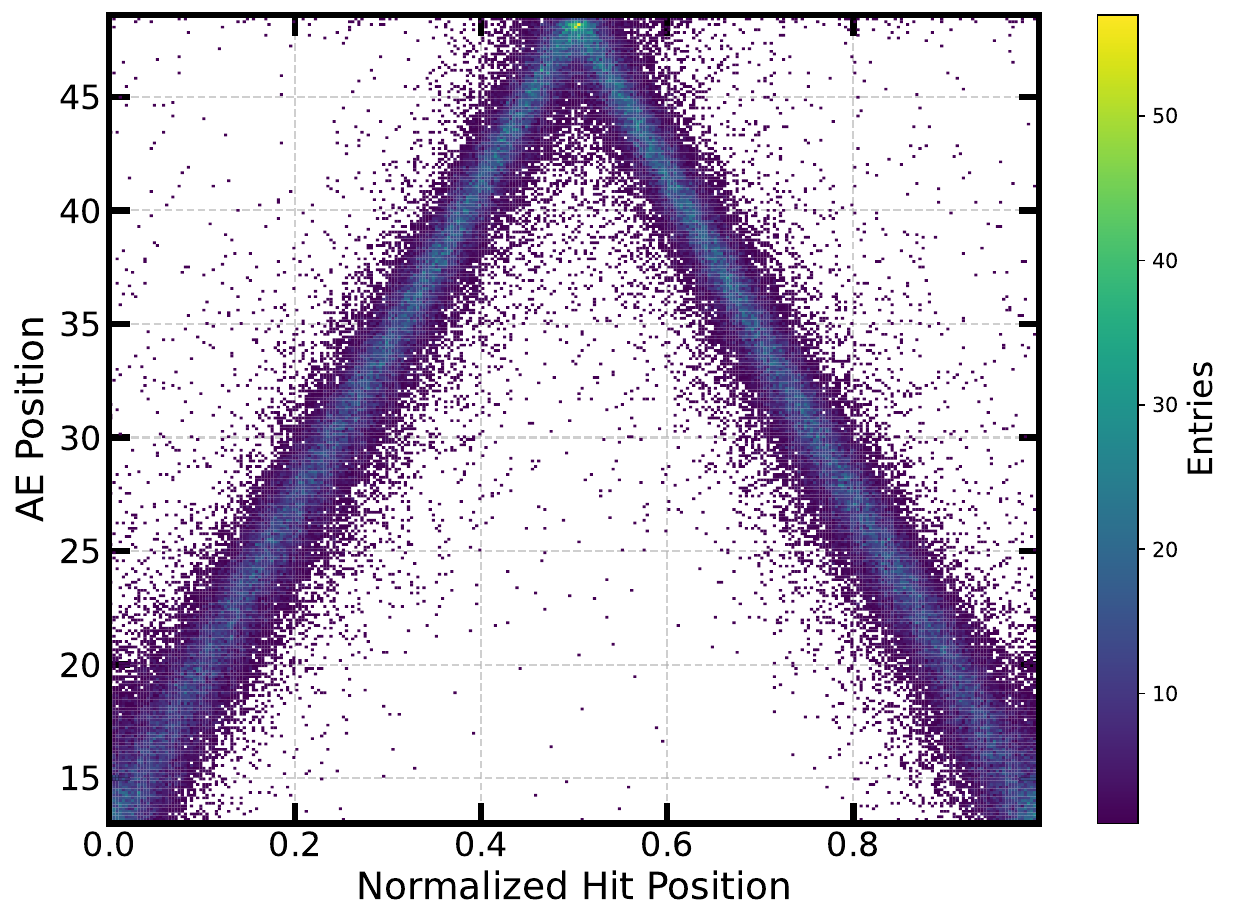}
        \caption{}
    \end{subfigure}
    \captionsetup{justification=raggedright}
    \caption{Physical interpretation of the HistoAE latent space. (a) Projection onto the charge dimension, where each point is color-coded by the independently measured charge from the telescope, shows that each AE-derived charge peak corresponds to the correct nuclei. (b) Comparison between the position dimension and the telescope-predicted impact position, normalized to the interval between two strips, reveals two distinct linear segments.
}
    \label{fig:result2}
\end{figure}

\textbf{(1) Precise control of the latent-space distribution.}
Compared with a plain AE or its variational and Wasserstein variants (VAE, WAE), the latent representation learned by HistoAE aligns exceptionally well with the physically motivated target distribution, namely a Gaussian mixture along the charge axis, direct product with a uniform distribution along the position axis. As shown in Fig.~\ref{fig:2dlatent} (a)(c), when applying previous solutions, such as the Wasserstein distance based loss, the latent variables indeed occupy an approximately square region in latent space, indicating that the support of the distribution is bounded as expected. However, the internal geometry within this region remains uncontrolled: the distribution exhibits irregular, curved band-like patterns with no clear physical meaning. While incorporating the HistoLoss, both the support and the detailed shape of the latent-space distribution align closely with the target, producing well-separated, linear structures as shown in Fig.~\ref{fig:2dlatent} (b)(d).

\textbf{(2) Fully interpretable latent space.}
Because the latent distribution learned by HistoAE matches the desired target manifold, the two latent dimensions acquire clear and independent physical interpretations. The dimension with a Gaussian-mixture structure corresponds to the particle charge, while the dimension with a uniform distribution represents the inter-strip position on the detector. Figure~\ref{fig:result2} shows the physical interpretation of the latent space along the two dimensions.
For the charge dimension, we color-code with the true charge of the incident particle. Each peak in Fig.~\ref{fig:result2} (a) corresponds to one charge species, ordered sequentially from low to high $Z$.
For the position dimension, we compare with the impact position predicted by the telescope in Fig.~\ref{fig:result2} (b), revealing a piecewise linear relation. This demonstrates that the latent variables jointly encode the complete physical information of the strip-signal pattern.

\begin{figure}[htbp]
    \centering
    \begin{subfigure}[b]{0.45\hsize}
        \includegraphics[width=\textwidth]{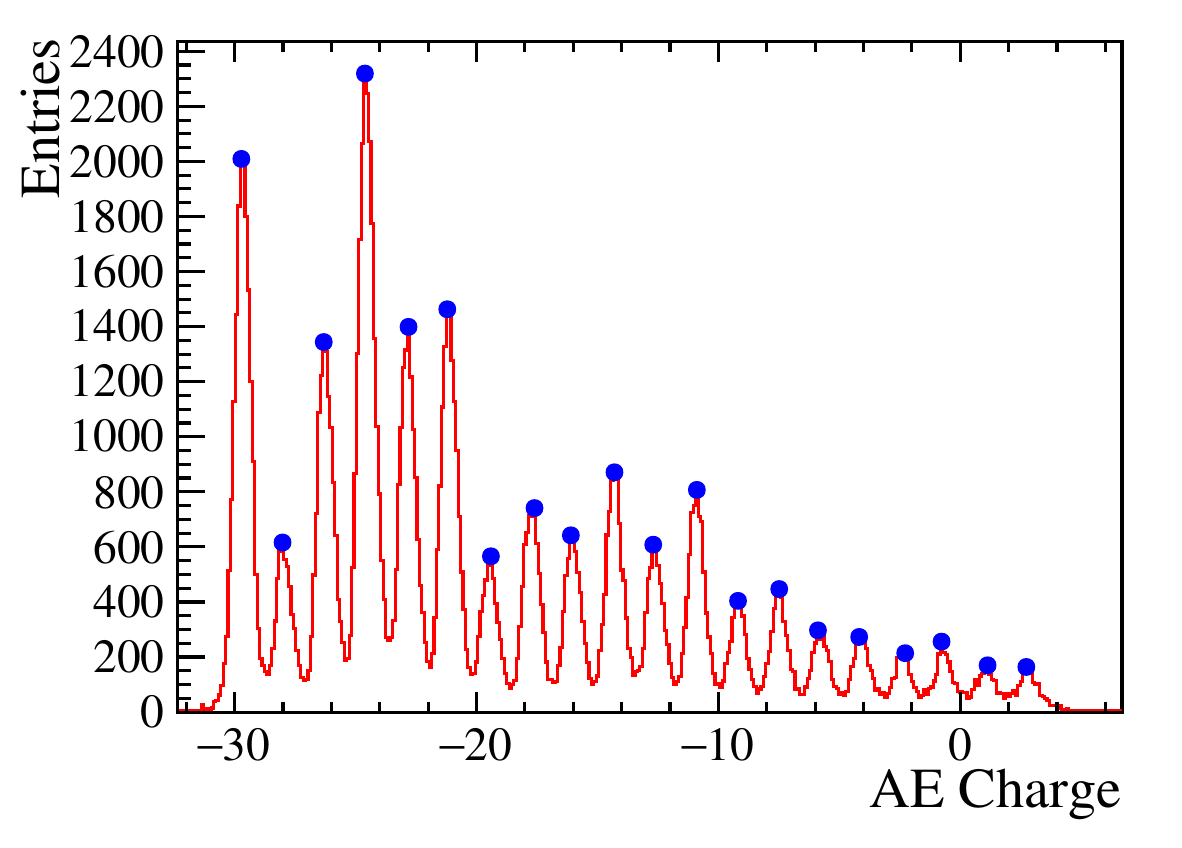}
        \caption{}
    \end{subfigure}
    \begin{subfigure}[b]{0.45\hsize}
        \includegraphics[width=\textwidth]{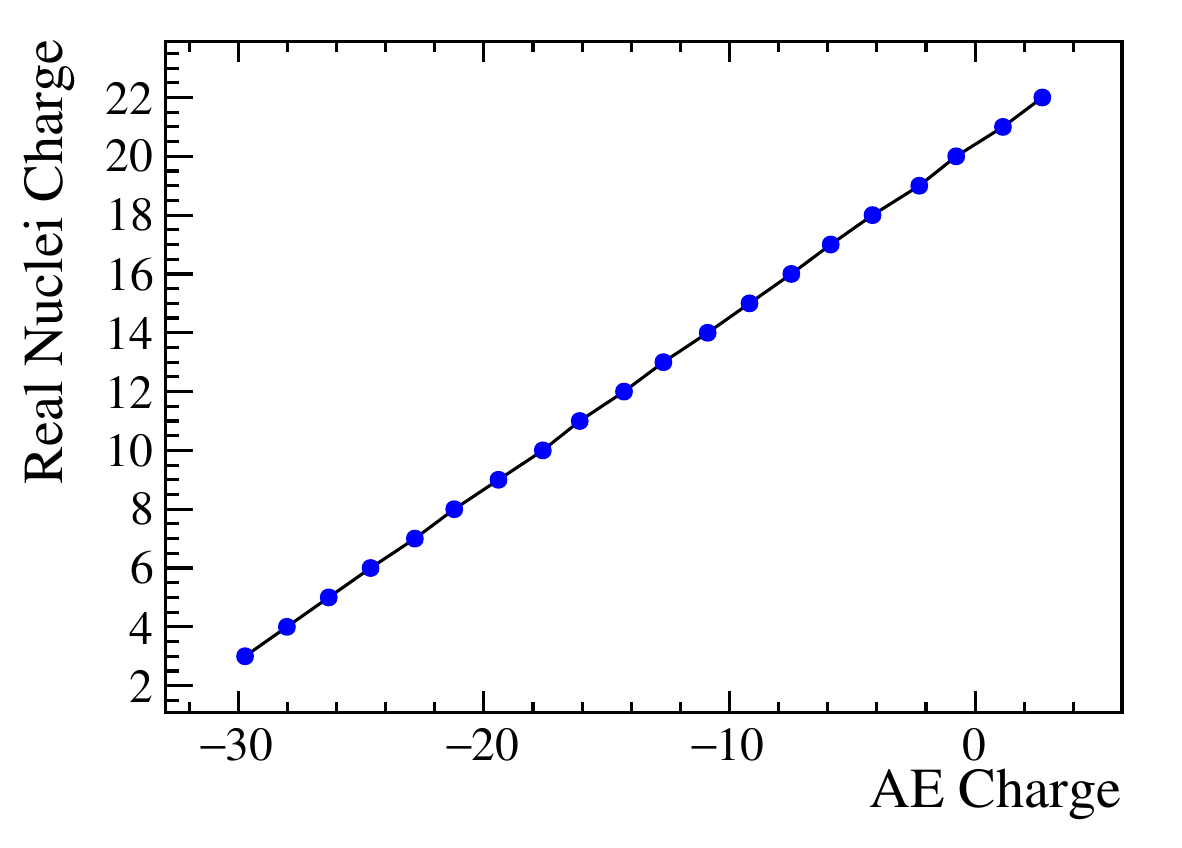}
        \caption{}
    \end{subfigure}

    \vskip 0.3cm

    \begin{subfigure}[b]{0.45\hsize}
        \includegraphics[width=\textwidth]{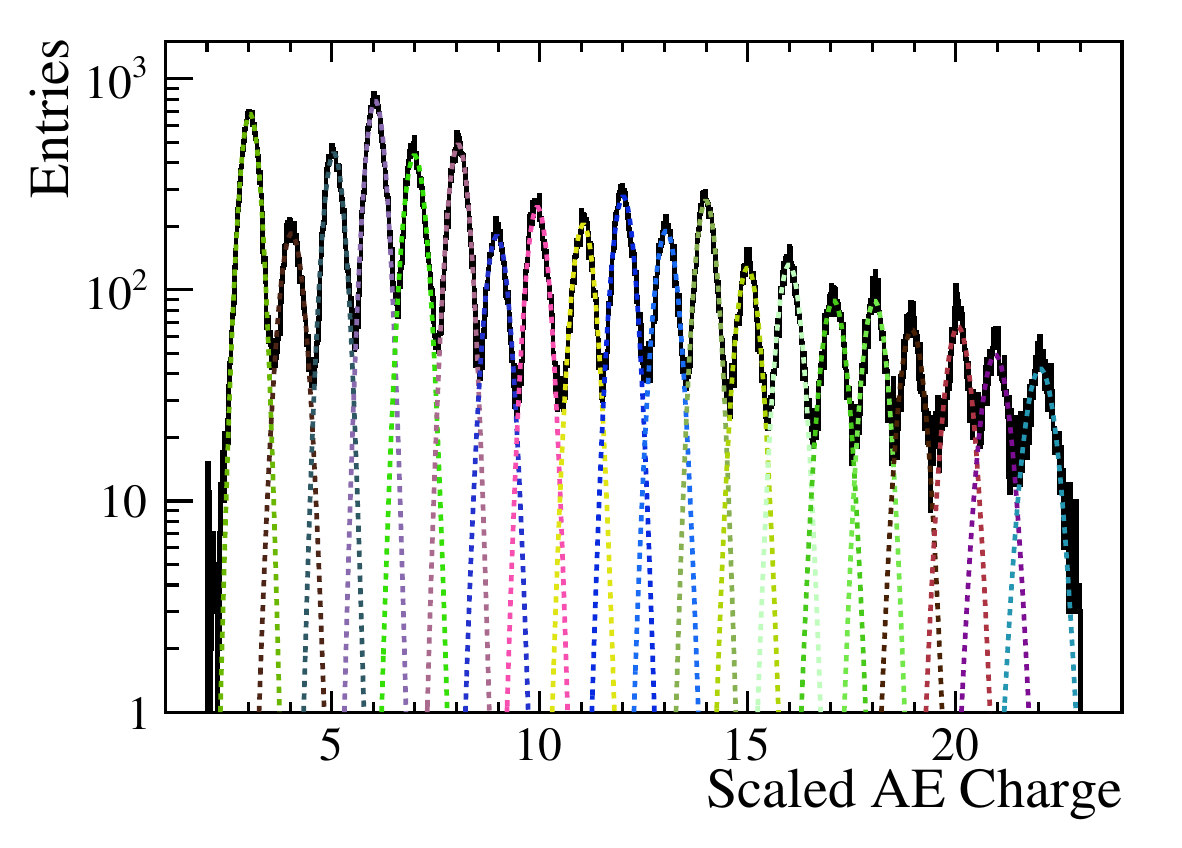}
        \caption{}
    \end{subfigure}
    \begin{subfigure}[b]{0.45\hsize}
        \includegraphics[width=\textwidth]{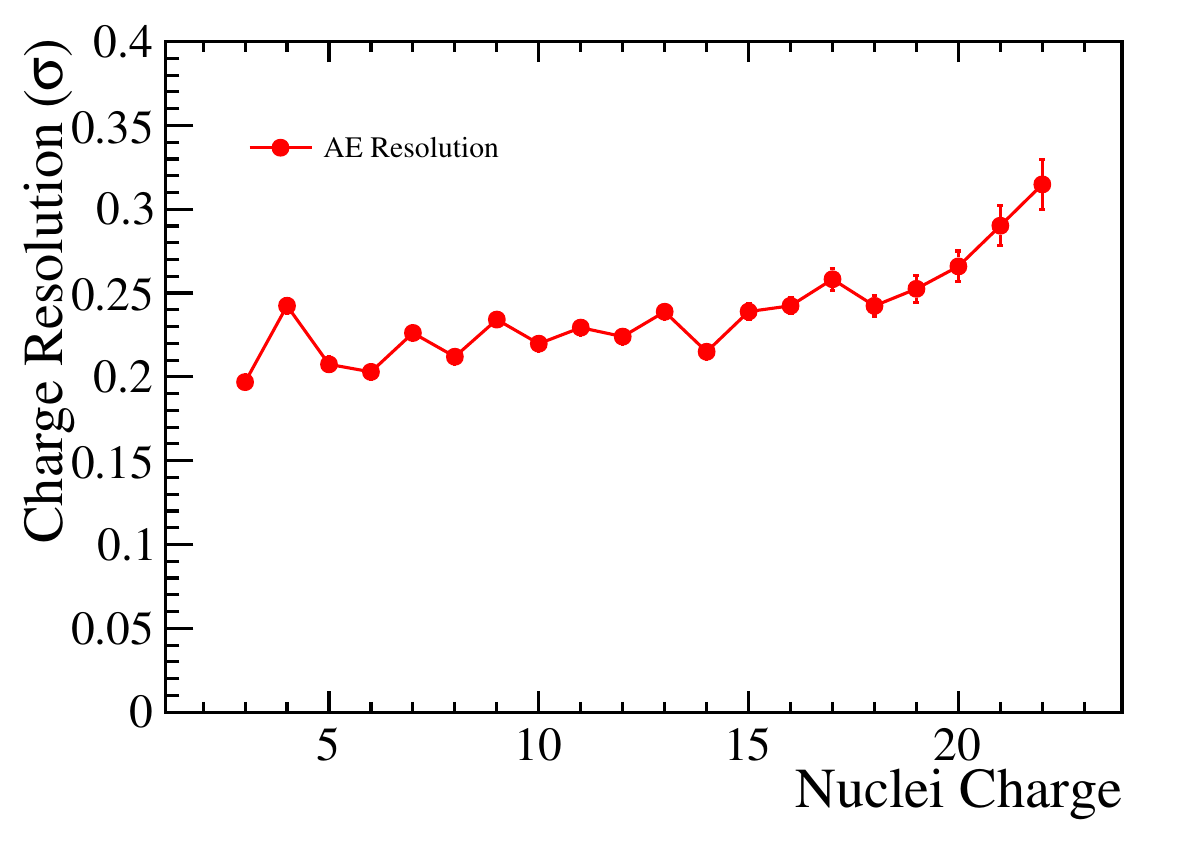}
        \caption{}
    \end{subfigure}

    \captionsetup{justification=raggedright}
    \caption{Charge measurement using HistoAE. (a) Local peak finding to determine the AE charge value corresponding to each integer charge nucleus. (b) Linear correlation between the AE charge and the true integer charge. (c) Rescaled AE charge to the physical charge space through interpolation based on the linear relation, and each nucleus peak is fitted with a Gaussian function. (d) The Gaussian width represents the charge resolution for each nucleus obtained with HistoAE.}
    \label{fig:charge_measure}
\end{figure}

\textbf{(3) Unsupervised precision measurement of physical quantities.}
With this interpretable latent distribution, we perform unsupervised measurement of physical quantities directly from the learned latent variables. Because the latent-space control in HistoLoss operates on normalized coordinates, a post-processing step is applied to map these to physical quantities. For the charge dimension, we identify the charge peaks and assign them to the corresponding integer charges ($Z=3$–$22$), and interpolate in between, as shown in Fig.~\ref{fig:charge_measure} (a)(b). The resulting charge distribution is shown in Fig.~\ref{fig:charge_measure} (c), where the individual peak is fitted with a Gaussian function. The sigma values of the Gaussian represent the charge resolution for the corresponding nuclei. The charge resolution of nuclei between \( Z = 3\, \mathrm{(Li)} \) and \( Z = 22 \, \mathrm{(Ti)} \) is shown in Fig.~\ref{fig:charge_measure} (d), which is better than $0.3 \, e$ for all nuclei. 
% which represents the best resolution of SSD in the world. 
\begin{figure}[htbp]
    \centering
    \begin{subfigure}[b]{0.47\hsize}
        \includegraphics[width=\textwidth]{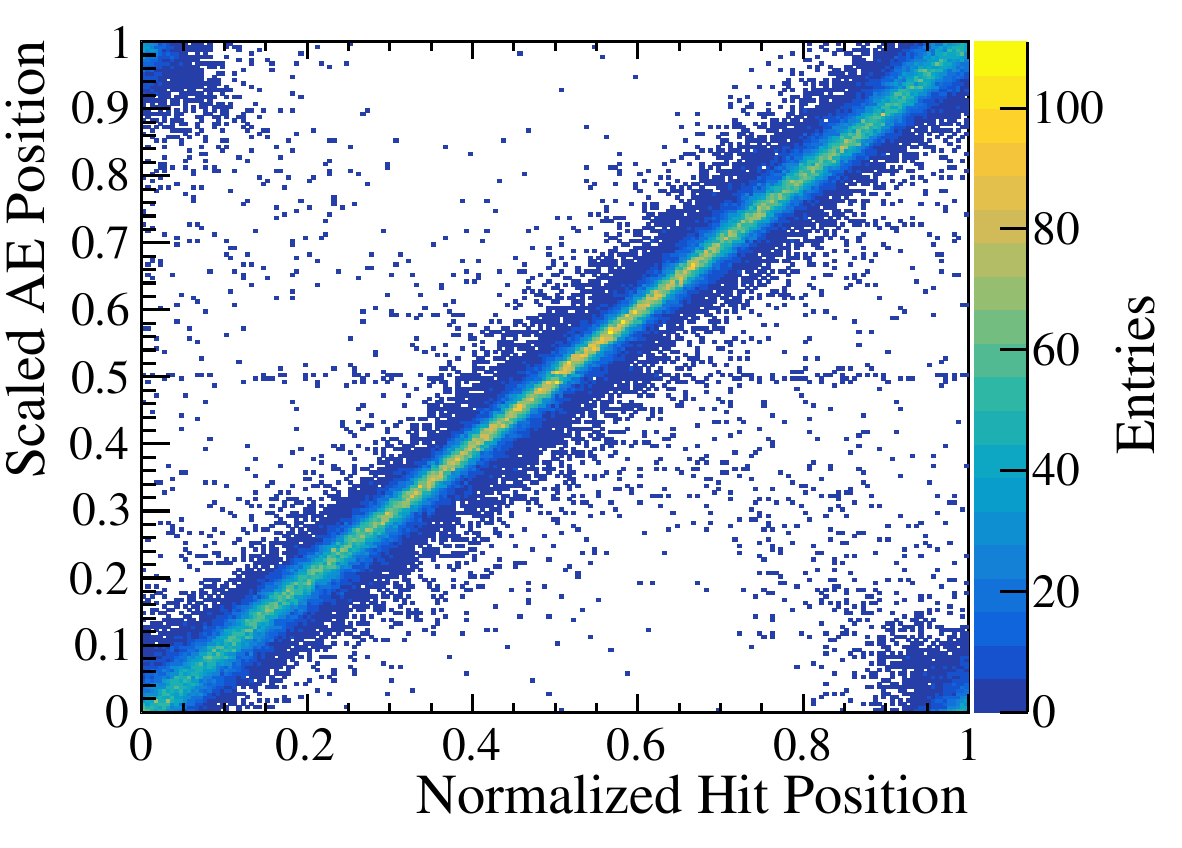}
        \caption{}
    \end{subfigure}
    \begin{subfigure}[b]{0.47\hsize}
        \includegraphics[width=\textwidth]{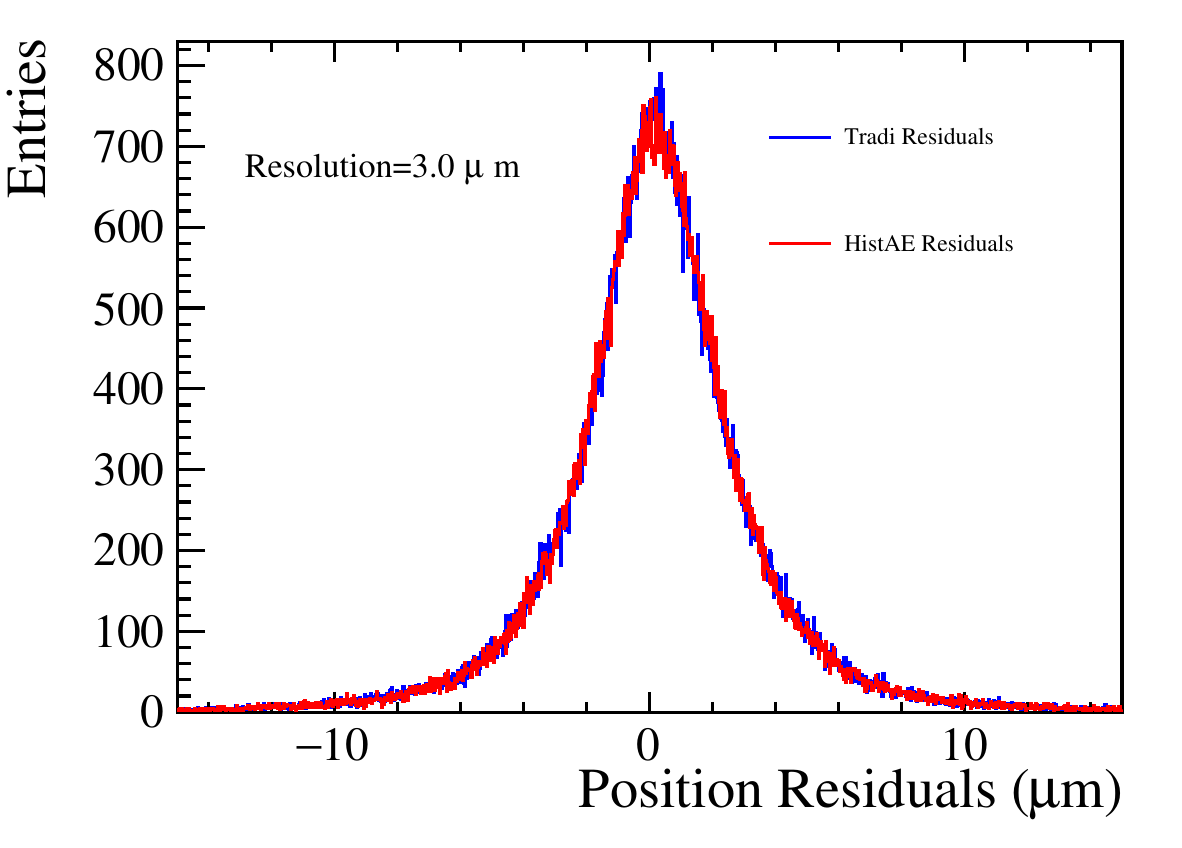}
        \caption{}
    \end{subfigure}
    \captionsetup{justification=raggedright}
    \caption{Hit position measurement using HistoAE. (a) Two-dimensional distribution of the scaled AE position versus the true impact position predicted by the telescope. (b) Residual distribution of the AE-reconstructed positions (red) and those obtained with the conventional method (blue). Both distributions are fitted with a double-Gaussian function, yielding a width of $3\,\mu$m.
 }
    \label{fig:position_measure}
\end{figure}
For position measurement, a normalization procedure is also required, in which the AE position is scaled to the range of 0–1. It can also be seen from Fig.~\ref{fig:result2} (b) that the AE position exhibits a piecewise linear relationship with the true impact position. This arises because particles hitting symmetrically between two strips produce mirror-symmetric clusters, while no left–right information is provided to the network. However, this ambiguity can be easily resolved by comparing the two largest channel signals within a cluster, denoted as ($S_L$) and ($S_R$), corresponding to the left and right signals, respectively. Define $\eta = \frac{S_R}{S_L + S_R}$. When 
$\eta < 0.5$, the impact point lies in the left half between the two strips, and vice versa. Therefore, we can express:
\[
\text{scaled AE position} =
\begin{cases}
\text{AE position}, & \eta < 0.5 \\
1 - \text{AE position}, & \eta \ge 0.5
\end{cases}
\]

As shown in Fig.~\ref{fig:position_measure} (a), the distribution of the scaled impact position versus the true position is presented. The residual distribution is fitted with a double-Gaussian function, from which the position resolution achieved by HistoAE is determined to be $3\,\mu$m, as shown in Fig.~\ref{fig:position_measure} (b). It can also be seen that the position resolution obtained by HistoAE is consistent with that derived from the conventional method, which has already reached the intrinsic performance limit of the detector~\cite{Turchetta:1993vu}. This indicates that HistoAE introduces no additional measurement uncertainty.

This demonstrates, for the first time, that highly precise measurements are achieved with a completely unsupervised learning method. Notably, HistoAE is also the first unsupervised algorithm capable of performing simultaneous charge and position reconstruction with SSDs, without requiring any labeled training data.

\begin{figure}[htbp]
    \centering
    \begin{subfigure}[b]{0.45\hsize}
        \includegraphics[width=\textwidth]{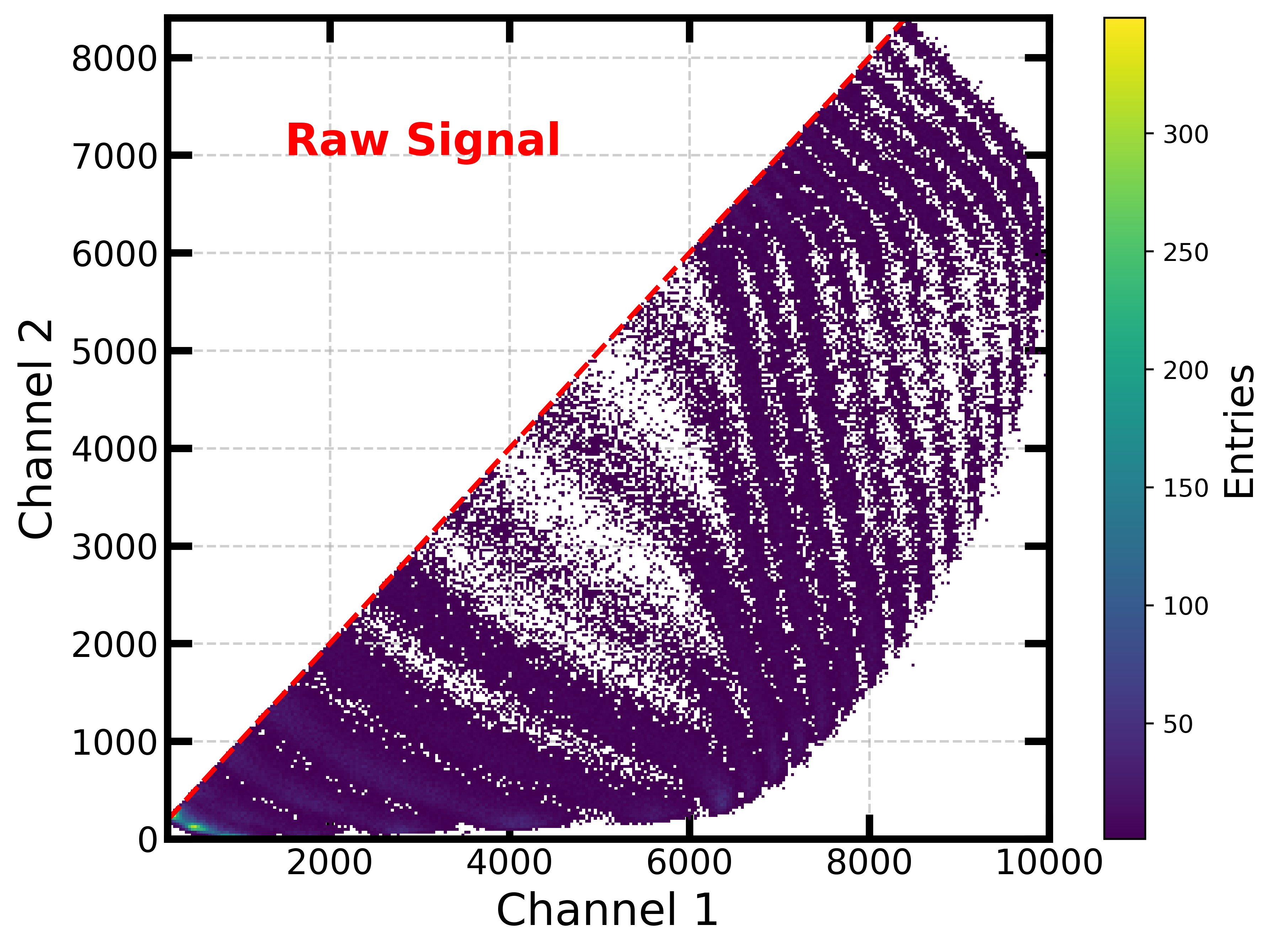}
        \caption{}
    \end{subfigure}
    \begin{subfigure}[b]{0.45\hsize}
        \includegraphics[width=\textwidth]{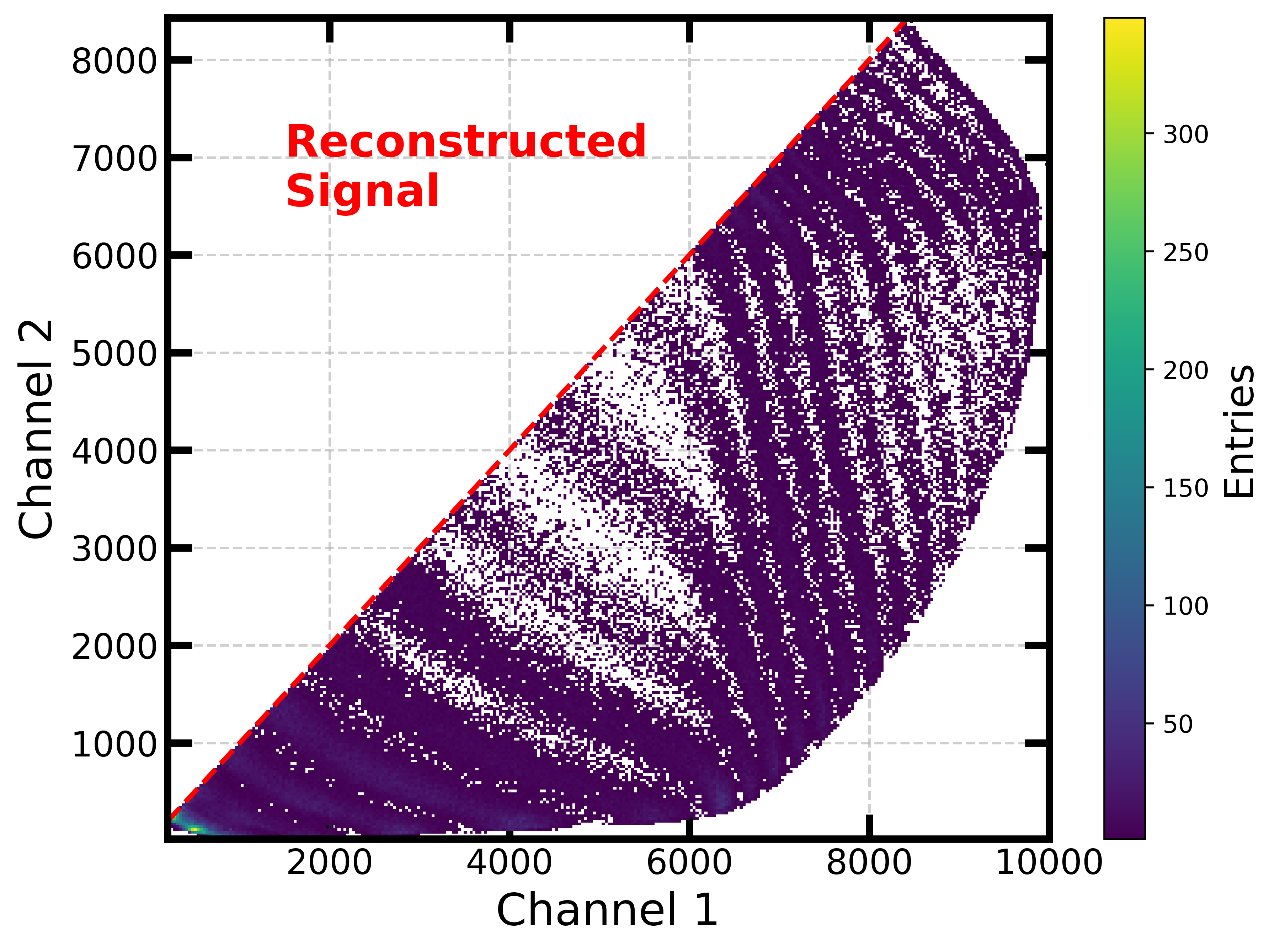}
        \caption{}
    \end{subfigure}
    \captionsetup{justification=raggedright}
    \caption{Comparison between the original and HistoAE reconstructed clusters, illustrated by the two-dimensional distribution of the two largest channel signals: (a) original cluster, (b) reconstructed cluster.
}
    \label{fig: recon}
\end{figure}

\textbf{(4) Fast simulation of SSD response.} Accurate and fast simulation of particle physics processes is crucial for the high-energy physics community~\cite{Apostolakis:2022nnf, Hashemi:2023rgo}. 
Regarding the use of HistoAE for fast simulation, this possibility is enabled by the fact that the learned latent-space distribution intrinsically captures the detector-response fluctuations, while the decoder maps these latent representations back to realistic detector clusters. The reconstruction quality of the decoder is illustrated directly in Fig.~\ref{fig: recon}, where the reconstructed cluster distribution reproduces the main structures of the original data. For example, for incident particles with charge $Z=10$, one may smear the charge coordinate according to the corresponding learned latent-space distribution and then pass the resulting latent representations through the decoder to generate detector clusters. In this way, the generated clusters naturally reproduce the characteristic band structure (just like the pattern shown in Fig.~\ref{fig:seed_eta}) associated with $Z=10$.

Therefore, the decoder of HistoAE can serve as a physically motivated generator of SSD responses. At the same time, for applications to downstream physics analyses, the quantitative accuracy of such a fast-simulation approach still needs to be evaluated in future studies.

\section{SUMMARY\label{sec conclusion}}
Unsupervised learning is widely used, yet its application to high-precision physical measurements has remained challenging. We present HistoAE, an unsupervised deep learning model capable of performing precise, interpretable measurements. When applied to silicon strip detectors, HistoAE learns a two-dimensional latent space whose axes correspond directly to particle charge and impact position. It achieves charge and position resolutions of approximately $0.25\,e$ and $3\,\mu$m, on par with conventional approaches. Moreover, its decoder can be repurposed for fast detector simulation. At present, the method is suitable for low-dimensional latent spaces. For tasks involving higher-dimensional latent representations, larger statistical samples are required to achieve effective control of the latent space. It will be addressed in future work.

Looking ahead, we also plan to further develop HistoAE for the future in-orbit operation of the AMS Layer-0 detector. In this context, HistoAE may offer two potential advantages: first, by reconstructing charge and position simultaneously in a unified representation-learning framework, it may help reduce the mutual error propagation present in conventional sequential corrections; second, because it does not rely on event-by-event labels from other subdetectors during training, it may help preserve more usable physics events in the large-acceptance L0 configuration. These possible advantages will need to be investigated carefully with real in-orbit data in future studies.

\acknowledgments
This study was supported by National Key Program for S\&T Research and Development (Grant NO.: 2022YFA1604800), and the National Natural Science Foundation of China (Grant NO.: 12342503). We would like to express our sincere gratitude to Luya Lou (AMSS), Shudong Wang (IHEP), Yiming Wang (IHEP), Zhiyu Xiang (CSU), Jiaoyang Xu (BNU), Yepeng Yan (IHEP), and Tongtian Zhu (ZJU) for the inspiring discussions that greatly facilitated the design of HistoAE. 
% We also wish to thank Professor Hung-yi Lee of National Taiwan University (NTU) for his open online lectures, which served as our foundational learning materials for machine learning.

\bibliography{apssamp}

%\nocite{*}
%
\end{document}